\documentclass[12pt,letterpaper,onecolumn]{article}
\usepackage{authblk}
\usepackage[latin1]{inputenc}
\usepackage{amsmath}
\usepackage{amsfonts}
\usepackage{amssymb}
\usepackage{makeidx}
\usepackage{graphicx}
\usepackage[left=2.50cm, right=2.50cm, top=2.50cm, bottom=2.50cm]{geometry}
\usepackage[section]{placeins}
\usepackage{float}

	\title{Ferro-lattice-distortions and charge fluctuations in superconducting LaO$_{1-x}$F$_{x}$BiS$_{2}$}
	
	\author[1]{A. Athauda}
    \author[2]{C. Hoffman}
	\author[3]{Y. Ren}
	\author[4]{S. Aswartham}
	\author[4]{J. Terzic}
	\author[4]{G. Cao}
	\author[4,5]{X. Zhu}
	\author[1]{D. Louca\thanks{Corresponding author, louca@virginia.edu}}
	\affil[1]{Department of Physics, University of Virginia, Charlottesville, VA 22904, USA.}
	\affil[2]{Oak Ridge National Laboratory, Oak Ridge, TN 37831, USA.}
	\affil[3]{Argonne National Laboratory, Lemont, IL 60439, USA.}	
	\affil[4]{Department of Physics and Astronomy, University of Kentucky, Lexington,KY 40506, USA.}	
	\affil[5]{Hefei High Magnetic Field Laboratory, Chinese Academy of Sciences, Hefei, China}	
\begin{document}

\maketitle

\begin{abstract}
	{Competing ferroelectric and charge density wave phases have been proposed to be present in the electron-phonon coupled LaO$_{1-x}$F$_{x}$BiS$_{2}$ superconductor.  The lattice instability arises from unstable phonon modes that can break the crystal symmetry. Upon examination of the crystal structure using single crystal diffraction, we find a superlattice pattern arising  from coherent in-plane displacements of the sulfur atoms in the BiS$_{2}$ superconducting planes. The distortions morph into coordinated ferro-distortive patterns, challenging previous symmetry suggestions including the possible presence of unstable antiferro-distortive patterns. The ferro-distortive pattern remains in the superconducting state, but with the displacements diminished in magnitude. Moreover, the sulfur displacements can exist in several polytypes stacked along the c-axis. Charge carriers can get trapped in the lattice deformations reducing the effective number of carriers available for pairing.}
\end{abstract}	

\newpage
A new class of phonon-mediated superconductors based on BiS$_{2}$ layers was
recently discovered \cite{1,2}. \ The structural features leading to a
proposed lattice instability are key to understand the electron-phonon
coupling mechanism, a central issue in the superconducting LaO$_{1-x}$F$_{x}%
$BiS$_{2}$ \cite{1,3,4,5} of interest in our study. \ The quasi
two-dimensional crystal structure with the superconducting BiS$_{2}$ bi-layers
sandwiched between insulating LaO layers (Fig. 1A) is reminiscent of the
cuprate \cite{6} and iron based superconductors. The Bi atoms form square pyramidal
units with in-plane sulfur S1 and apical S2 atoms. \ First-principles, band
structure and spin polarization calculations indicated that a phonon
instability breaks the nominal tetragonal symmetry, \textit{P4/nmm}, where the
atomic distortions are described using symmetries such as \textit{P2}$_{1}%
$\textit{mn }\cite{4}, non-centrosymmetric \textit{C}$_{2}$ or the
centrosymmetric \textit{P2}$_{1}$\textit{/m }\cite{7}. \ Such a change in the
crystal symmetry will have significant implications on the band structure thus
an experimental verification is necessary.

The Fermi surface is dominated by the hybridized orbitals of Bi $6p$ and S
$3p$. \ Upon fluorine doping via the substitution of oxygen in the insulating
layer, as in LaO$_{^{1-x}}$F$_{x}$BiS$_{2}$, electron carriers are introduced
in the lattice \cite{8} and the density of states at the Fermi level
increases \cite{3} while the nominal symmetry remains unchanged. Theoretical
works \cite{4,5,9} suggested that a large phonon softening occurs due to
dynamic sulfur displacements, leading to an instability around the $\Gamma$
point in x = 0, and along Q = ($\pi$,$\pi$,0) with doping, that has been
associated with strong Fermi surface nesting. Earlier neutron powder
diffraction measurements showed that indeed sulfur is displaced from its
equilibrium position \cite{10}, consistent with theoretical predictions, and
can possibly lead to charge fluctuations, similar to the parent compound of
another phonon superconductor, BaBiO$_{3}$ \cite{11}. However, to obtain the
three-dimensional arrangement of the atomic distortions and possible symmetry
breaking conditions, high-resolution single crystal diffraction is clearly
needed. \ To this end, synchrotron X-ray and neutron experiments were carried
out on single crystals of LaOBiS$_{2}$ and superconducting LaO$_{0.7}$%
F$_{0.3}$BiS$_{2}$ with a T$_{C}$ of $\sim$2.5 K.

The single crystals were grown using the CsCl/KCl flux method and SEM/EDX was
used to determine their stoichiometry, while the bulk magnetization
measurements were carried out using a SQUID magnetometer. \ The synchrotron
X-ray experiments on LaOBiS$_{2}$ and LaO$_{0.7}$F$_{0.3}$BiS$_{2}$ were
performed at 11-ID-C using X-rays of 105 KeV at the Advanced Photon Source at
Argonne National Laboratory. The data were collected at a wavelength of
0.11798 \AA \ in transmission mode. \ The neutron scattering experiments were
carried out at the single crystal diffractometer TOPAZ at Oak Ridge National
Laboratory. The neutron data were background subtracted and corrected for
absorption, neutron path length, spectrum, detector efficiency and Lorentz
factor. The LaO$_{0.7}$F$_{0.3}$BiS$_{2}$ crystal showed single Bragg peaks at
each \textit{hkl} site. The parent compound however showed several Bragg spots
at each \textit{hkl} site indicating several grains. In this case, attention
was paid to include in the refinement Bragg peaks from the same grain only by
calculating the diffraction condition at each Bragg point. The integrated
Bragg intensities were determined using the 3D spherical Q space integration and
data refinement were performed using the ShelX software \cite{12}.

The BiS$_{2}$ planes are buckled with equal bond lengths between Bi and S1 at
2.874 \AA \ in the \textit{P4/nmm symmetry }\cite{2,13,14,15,16,17} while the
bonding between the Bi ion and the apical S2 is shorter, at 2.466 \AA , as
shown in Fig. 1A. \ The neutron diffraction pattern shown in Fig. 1B with
the Bragg spots corresponding to \textit{P4/nmm }labeled is from the
LaOBiS$_{2}$ single crystal. \ It can readily be seen that several Bragg spots
(within the circle) are not reproduced by this symmetry. \ Shown in Fig. 1C
is the calculated pattern in color dots from \textit{P4/nmm }that based on its
reflection condition in the hk0 plane (h+k = 2n), odd combinations of h+k are
forbidden. \ This is consistently observed in the superconducting crystal as
well, as will be shown below. \ The integrated intensity\ of the neutron data
across the \textit{h}$\overline{3}$\textit{0} line is plotted in Fig. 1D.
\ The forbidden reflections are much weaker by comparison to the main Bragg
peaks. \ What is the underlying symmetry that can yield the right diffraction
pattern? 

To answer this question, we test previously proposed symmetries. \ The
calculated diffraction patterns from four different symmetries are shown in
Fig. 2. \ The corresponding unit cells are shown in the supporting Tables.
\ \textit{P4mm} and \textit{P}$\overline{4}$\textit{m2 }\cite{8} are
subgroups of \textit{P4/nmm} (Fig. 2C and 2D). \ Given that the same
reflection conditions apply as in the \textit{P4/nmm}, neither symmetry can
reproduce the new reflection conditions. \ In the \textit{P2}$_{1}$\textit{mn
}(Fig. 2B) symmetry suggested in Ref. \cite{4}, not all reflections can be
reproduced by the calculated pattern either, because the reflection condition
along the \textit{h00} direction of h = 2n restricts the symmetry from
producing extra reflections beyond the ones obtained in the
\textit{P4/nmm} symmetry. \ A final comparison with a fourth symmetry,
\textit{P2}$_{1}$\textit{/m}, yields a Bragg structure with extra reflections
as seen in Fig. 2A, but fails to reproduce critical reflections at spots marked in
the figure. \ This leads us to conclude that the symmetry of LaO$_{1-x}$%
F$_{x}$BiS$_{2}$ is considerably lower than expected.

Of significance to the crystal structure is the motion of S1. \ Earlier
results hinted at in-plane sulfur atom, S1, displacements possibly
leading to ferroelectric like modes \cite{4}, although no structural transition was
expected because of quantum zero-point motions that render the system
dynamically disordered. Our earlier neutron powder diffraction measurements
yielded a similar outcome but with a larger amplitude of displacement (about 0.3 \AA \ \cite{10}). Several displacement patterns can exist that can be either
ferro-or antiferro-distortive in nature, in which the two S1 atoms in the
BiS$_{2}$ bi-layers are displaced simultaneously in the x- or y-directions,
either in parallel or anti-parallel. \ Distinction between the two patterns
cannot be made from the powder data as each model results in the same
magnitude of bonds regardless of orientation in real-space \cite{10}. 

Within one unit cell, three unique directions of the S1 displacements are
possible as listed here: (+x, +x), (+x, +y) and (+x, -x) where the two
coordinates refer to the S1 ion in the bi-layers. Given that the lattice is
nominally tetragonal and the a- and b-axes are equivalent, the magnitude of
the displacement is the same in both directions, and the direction, either -x-
or -y-, of the displaced ion can be altered.  However, the displacement
direction at one layer is important relative to the second layer. If the
displacements are antiferro-distortive, the antiferro-distortive unit cell is
doubled along the a- and b-axes (Fig. 3A). On the other hand, if the
displacements are ferro-distortive, the ferro-distortive cell is not doubled
(also shown in Fig. 3A). From the structure factor calculations, it is
determined that when the displacement pattern is (+x, -x), no intensity is
produced at Miller index h=0 when k is odd. When the displacement pattern is
(x, x), the intensity is zero again when either h or k become odd. Thus the
(+x, -x) and (+x, +x) patterns are excluded and not considered further.

Using the remaining pattern (+x, +y) (or equivalently (-x, -y)), the Bragg
reflections in the \textit{hk0} plane are calculated using the ferro- and
antiferro-distortive modes in the \textit{P1} symmetry. This pattern breaks
the fourfold, twofold and inversion symmetry operations of the nominal
\textit{P4/nmm}. In Fig. 3B and 3C, the calculated reciprocal lattice
patterns corresponding to the two modes are shown. Far more reflections are
observed in the antiferro-distortive mode, and can readily be eliminated when
compared to the data from LaOBiS$_{2}$ (Fig. 1B) since it generates half
integer peaks that are not actually present. On the other hand, the
ferro-distortive mode can reproduce all Bragg spots. \ Thus the ferro-distortive mode remains as the only possibility.

To reproduce the Bragg peak intensity, a full profile refinement of the three
dimensional neutron diffraction intensity is performed. This involves a total
of about 4600 Bragg peaks. Several polytypes with different stacking
distortion patterns along the c-axis are assumed. Four of them are shown in
Fig. 4A. Each quadrant in a domain corresponds to a BiS$_{2}$ plane. Thus
four quadrants correspond to four planes stacked along the c-axis where the
layers are labeled as 1, 2, 3, and 4. Thus each domain represents an S1
distortion pattern in four consecutive planes. The displacement vector of the
same magnitude designated by x or y are indicated with the arrows. The
displacement pattern for the first domain is (-x, -y, -x, -y), for the second
is (+x, +y, -x, -y), for the third is (-x, +y, -x, -y) and for the fourth is
(-x, +x, -y, +y). The calculated structure factor for each displacement mode
is compared to the observed structure function. All patterns produce a
comparable agreement as determined from their $\chi^{2}$ values (0.1414, 0.1373, 0.1408, 0.1383). At the same time, we find that when the
calculated intensity from all Bragg peaks is averaged over all four domain
structures, the fitting improves further and the result is shown in Fig. 3B
with the $\chi^{2}$. The plot between the observed and calculated squared
intensities is on the expected line (blue line with a slope of 1), where the
superlattice structure is well reproduced. \ Some differences are observed
involving the highest intensity Bragg peaks. This suggests that all four domains
are most likely present. Moreover, other displacement patterns are also
likely, with different sequencing patterns, thus the four domain pattern
presented here is just one solution to reproducing the peak intensity. \ It is
nonetheless clear that the sulfur ferro-displacement is the key component to
the broken crystal symmetry that can be significant in theoretical calculations.

To test the applicability of the model, the ferro-distortive cell is used to
fit the data collected from the 30 \% superconducting sample. Shown in Fig.
5A is the plot of the bulk magnetic susceptibility that shows a T$_{C}$
$\sim$2.5 K. \ The single crystals are not exposed to high pressure annealing
as in the case of powders hence their T$_{C}$'s are lower \cite{2,15}. At the
same time, the stacking faults observed in the powder samples that are induced
by the high-pressure treatment leading to a pronounced broadening of Bragg
peaks with an \textit{l}-component\cite{14} are absent in single crystals. In
Figs. 5B and 5C, it can readily be seen that the ferro-distortive mode can
reproduce all Bragg reflections, just like in the parent compound. However,
the magnitude of the x- and y-displacements of S1 is smaller, at 0.2 \AA . A
real-space arrangement of the four domains in the ab-plane can result in
regions with varying degree of spatial uniformity that may not be conducive to
superconductivity. When the domains are arranged along the c-axis, the
experimental diffraction pattern is reproduced well which is consistent with
the lattice accommodating many polytypes. \ However, the arrangement of the
domains in the ab-plane is more critical because domain walls can be created
which are antiferro-distortive. \ These can give rise to additional peaks not
present in the data as can be seen from the diffraction pattern in Fig. 5E
of the two-dimensional arrangement shown in Fig. 5D.  Thus, lattice
uniformity is important in the ab-plane,
but the polytypes can alternate along the c-axis. Charge carriers can get
trapped in the lattice deformations reducing the effective number of carriers
available for pairing.

The electronic structure and Fermi surface of LaOBiS$_{2}$ resemble that of
superconducting MgB$_{2}$ although T$_{C}$ is much higher in the latter
\cite{18,19,20}. Even though their coupling constants are comparable, T$_{C}$ is
significantly lower in the current system because the main contribution stems
from low frequency phonon modes \cite{8}. Can the proposed displacement mode
lead to charge disproportionation, ferroelectricity in Bi or a charge density
wave (CDW)? In the BaBiO$_{3}$ phonon superconductor, the Bi-O bonds are split
at 2.11 and 2.29 \AA \ \cite{21} due to a breathing \cite{11} or a
Peierls-like\cite{22} instability. However, with K doping as in Ba$_{0.6}%
$K$_{0.4}$BiO$_{3}$, the split disappears and the bond lengths become one,
2.14 \AA \ \cite{23,24}, which is not the case in the superconducting
LaO$_{0.7}$F$_{0.3}$BiS$_{2}$ in which the distortions are present, albeit reduced in magnitude.  In conclusion, the structure of
LaO$_{1-x}$F$_{x}$BiS$_{2}$ is exposed. The sulfur mode that can explain the single
crystal results is ferro-distortive in nature, and must be responsible for the
large electron-phonon coupling. This mode must bring significant changes in
the electronic structure around the Fermi level. Fluctuations of the bond
lengths in the superconducting crystal as shown here indicate that charge
fluctuations persist and are most likely significant in the mechanism of
superconductivity. Thus the electronic pairing and the form of
superconductivity present in this system is sensitive to the precise nature of
the crystal structure.\\

\begin{flushleft}
	{\large \textbf{Acknowledgments}}\\
\end{flushleft}

The authors would like to acknowledge support from the National Science
Foundation, DMR-1404994, 1265162 and 1334170. Work at ORNL was supported by
the US Department of Energy, Office of Basic Energy Sciences, Materials
Sciences and Engineering Division and Scientific User Facilities Division.
Work at the Advanced Photon Source, was supported by the US Department of
Energy, Office of Science User Facility, and operated by Argonne National
Laboratory under Contract No. DE-AC02-06CH11357.

\section*{}
	\begin{figure*}[ht]
		\begin{center}
			\centerline{\includegraphics[width=1\textwidth]{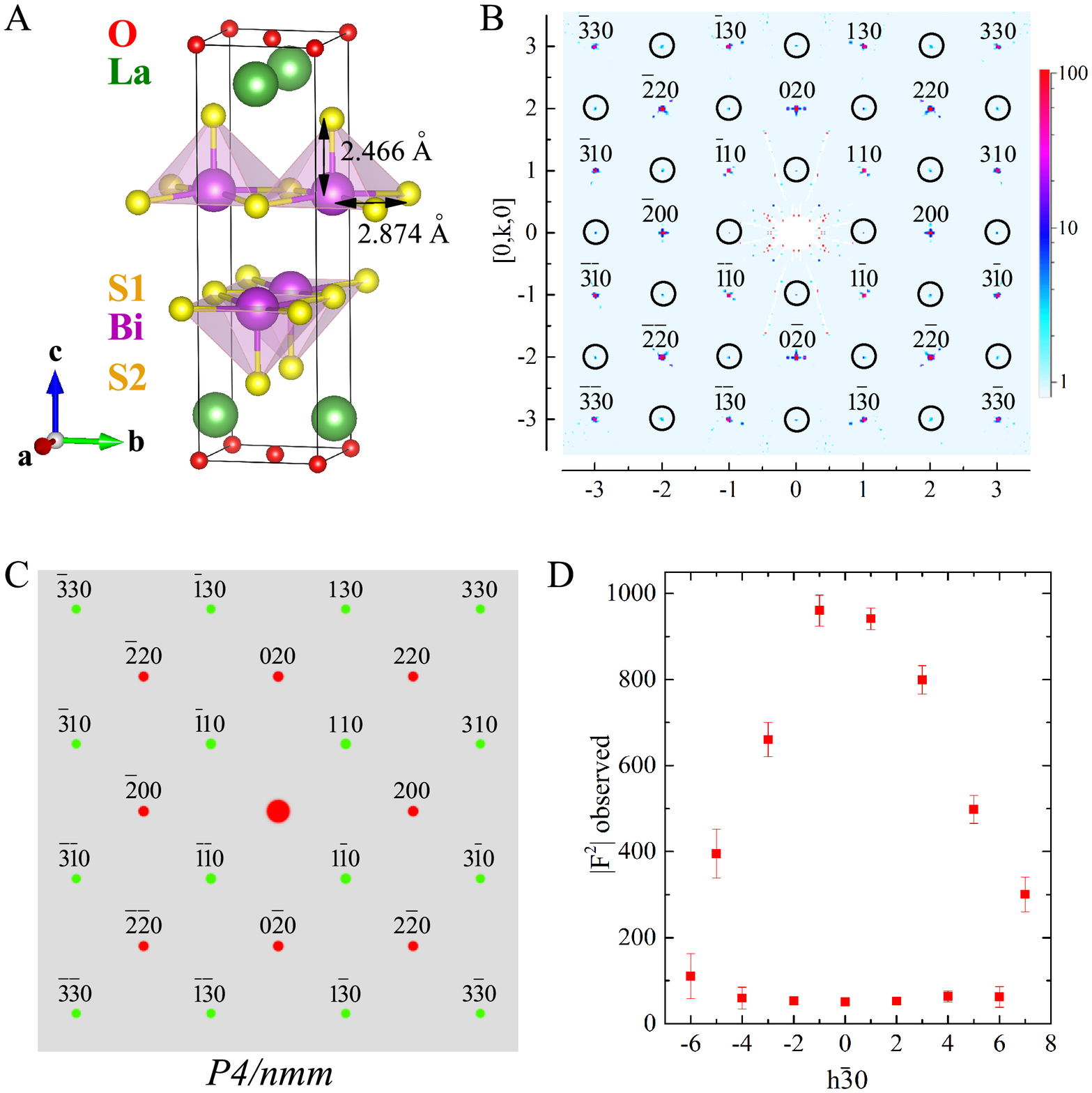}}
			\caption{In A, the layer crystal structure with the Bi-bilayer is shown. In B,
				the single crystal neutron diffraction pattern (black dots) in the hk0
				reciprocal plane for the parent compound, LaOBiS2, is shown. \ Several
				additional reflections are observed as circled. \ In C, the \textit{hk0} plane is constructed using the \textit{P4/nmm} symmetry.\ \ In D, the integrated intensity of the observed peaks is plotted along the h$\overline{3}$0 line.}\label{Figure1}
		\end{center}
	\end{figure*}
	
	\begin{figure*}[ht]
		\begin{center}
			\centerline{\includegraphics[width=1\textwidth]{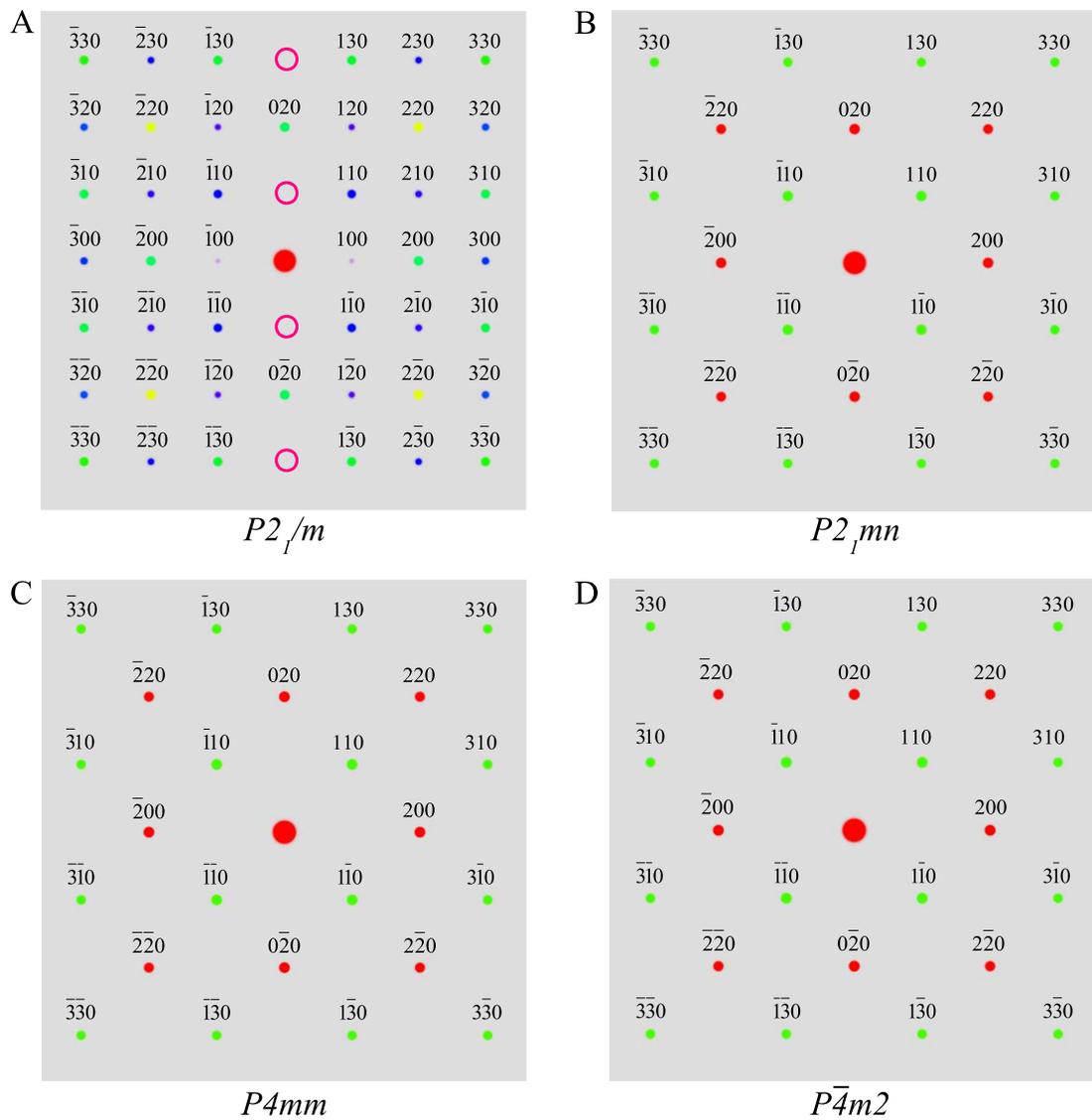}}
			\caption{In A through D, the calculated reciprocal lattice patterns based on
				the four symmetries.}\label{Figure2}
		\end{center}
	\end{figure*}
	
	\begin{figure*}[ht]
		\begin{center}
			\centerline{\includegraphics[width=1\textwidth]{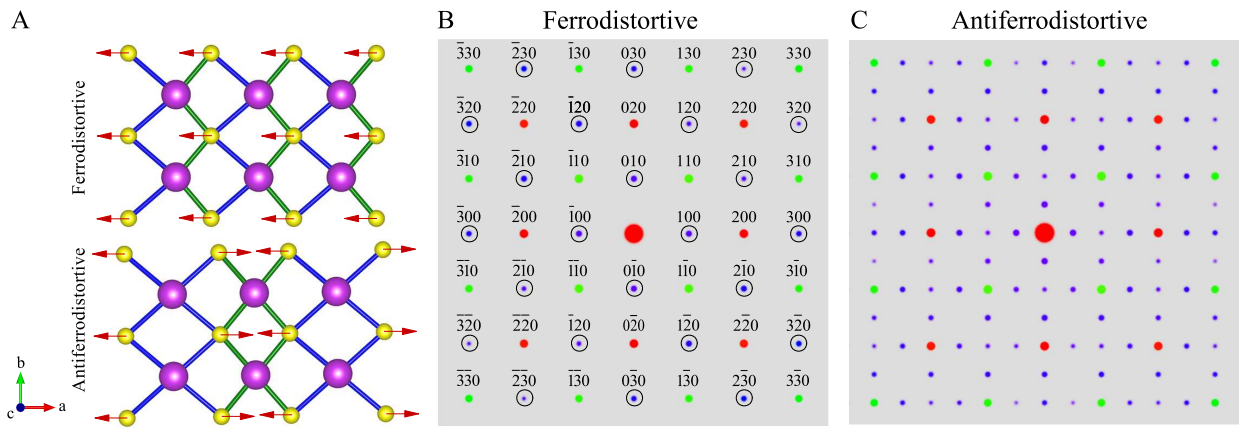}}
			\caption{The ferro-distortive and anti-ferrodistortive displacements modes are
				shown in A. \ Only in-plane S1 atoms are allowed to move. In B, the calculated diffraction pattern from the ferro-distortive mode is shown while in C, the diffraction pattern resulting from the antiferro-distortive mode is shown.}\label{Figure3}
		\end{center}
	\end{figure*}
	
	\begin{figure*}[ht]
		\begin{center}
			\centerline{\includegraphics[width=1\textwidth]{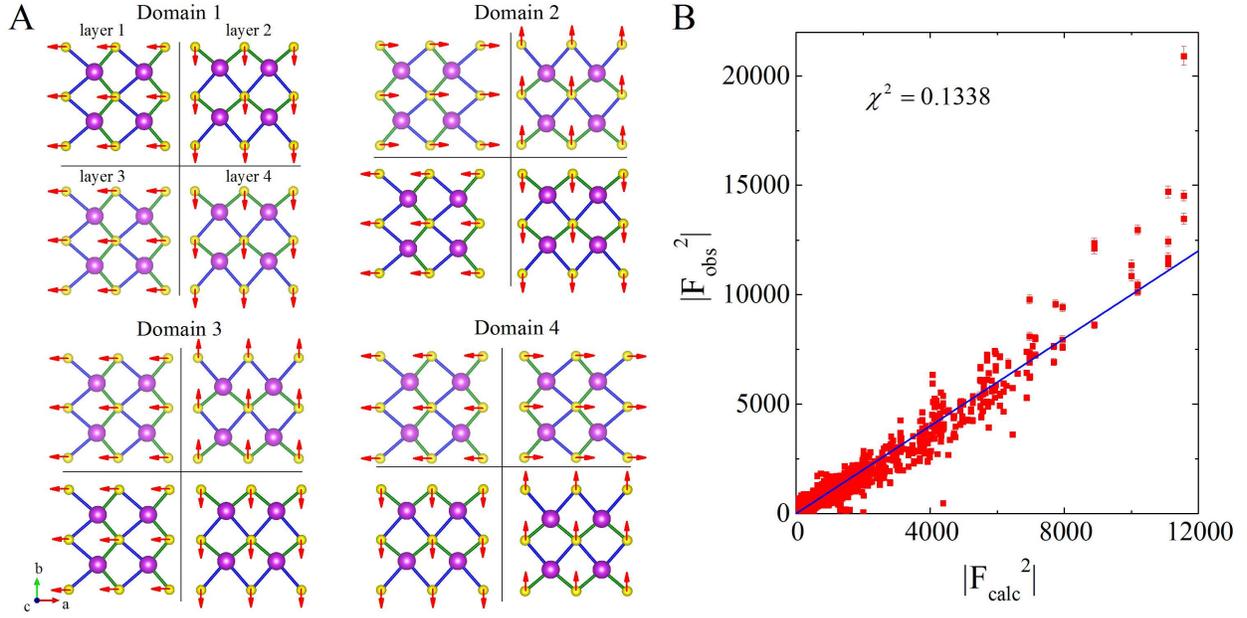}}
			\caption{In A, four domain structures corresponding to displacements of S1
				along the \textit{c}-axis are shown. Each domain contains four quadrants where
				each quadrant represents one BiS$_{2}$ layer along the \textit{c}-axis. In B,
				the square of the structure function,
				$\vert$%
				F(Q)%
				$\vert$%
				2, for the observed intensity is compared to the calculated intensity obtained
				from the refinement based on averaging over all four domains. The fitting is
				done using the neutron single crystal data. The goodness of fit is determined
				to be 0.133. The blue line represents the ideal match between the calculated
				and the observed value at each Bragg peak with a slope of 1. Deviations are
				observed for the highest intensity peaks.}\label{Figure4}
		\end{center}
	\end{figure*}
	
	\begin{figure*}[ht]
		\begin{center}
			\centerline{\includegraphics[width=1\textwidth]{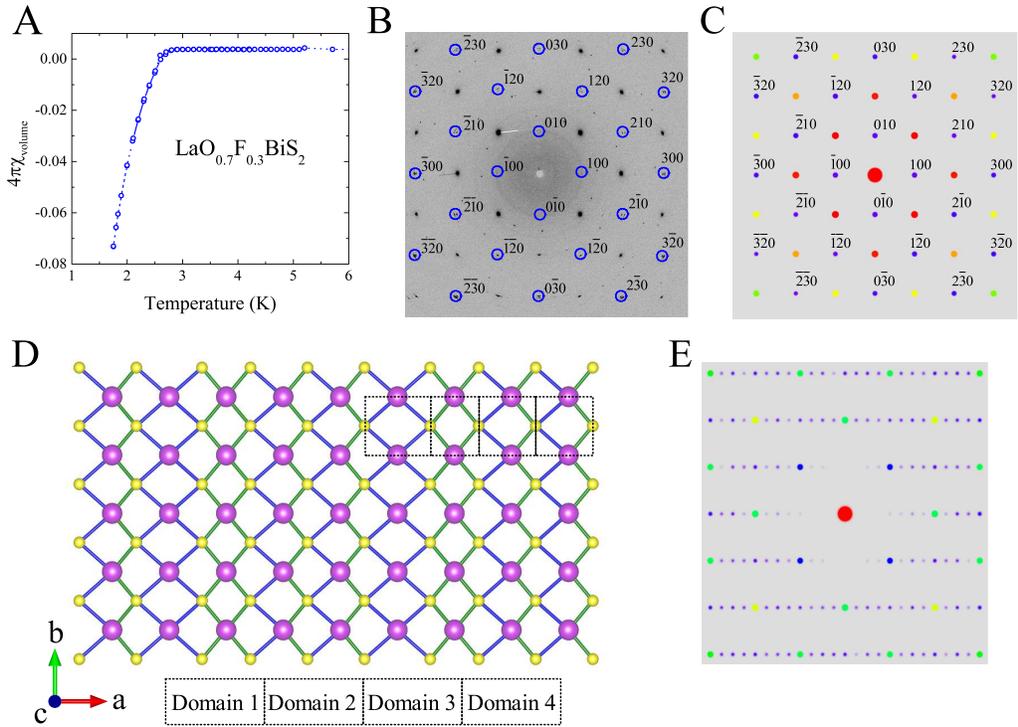}}
			\caption{In A, the bulk magnetic susceptibility of the x = 0.3 crystal is
				shown. In B and C, the X-ray data in the \textit{hk0} plane and the calculated
				pattern assuming the ferro-distortive pattern are shown. In D, an example is
				shown in which the domains are arranged in the ab-plane. Domain walls appear
				at the interfaces when the distortions change orientation and give rise to
				antiferro-distortive displacements. Shown in E is the calculated pattern from
				the real-space arrangement shown in D. Extra reflections are produced from
				this model that do not match the pattern shown in B.}\label{Figure5}
		\end{center}
	\end{figure*}
	
\newpage

\section*{Supporting Information}

\begin{table}[ht]
	\centering
	\caption{The tested model of P2$_{1}$/m [25] symmetry }
	\label{Table S1}
\begin{tabular}{|cccc|}
	\hline
	\multicolumn{2}{|c}{LaOBiS$_{2}$} & \multicolumn{2}{c|}{P2$_{1}$/m (\#11)} \\ \hline
	& \textit{a}(\AA{})            & 4.0769          &                 \\
	& \textit{b}(\AA{})            & 4.0618          &                 \\
	& \textit{c}(\AA{})            & 13.885          &                 \\
	& $\beta$(°)            & 90.12           &                 \\
	& & & \\
	            & x & y & z\\
	La          & 0.2496          & 0.25            & 0.0899          \\
	Bi          & 0.2346          & 0.25            & 0.6309          \\
	S1          & 0.219           & 0.25            & 0.3844          \\
	S2          & 0.2483          & 0.25            & 0.8094          \\
	O           & 0.753           & 0.25            & 0.002           \\ \hline
\end{tabular}
\end{table}

\begin{table}[ht]
	\centering
	\caption{The tested model of P2$_{1}$mn [4] symmetry}
	\label{Table S2}
\begin{tabular}{|cccc|}
	\hline
	\multicolumn{2}{|c}{LaOBiS$_{2}$} & \multicolumn{2}{c|}{P2$_{1}$mn (\#31)} \\ \hline
	& \textit{a}(\AA{})            & 4.037           &                 \\
	& \textit{b}(\AA{})            & 4.029           &                 \\
	& \textit{c}(\AA{})            & 14.217          &                 \\
		& & & \\
		& x & y & z\\
		La         & 0.4997         & 0             & 0.0858           \\
		Bi         & 0.4973         & 0             & 0.6324           \\
		S1         & 0.525          & 0             & 0.3946           \\
		S2         & 0.4988         & 0             & 0.8102           \\
		O          & 0.4996         & 0.5           & 0    \\
		\hline           
	\end{tabular}
\end{table}

\begin{table}[ht]
	\centering
	\caption{The tested model of P4mm symmetry}
	\label{Table S3}
	\begin{tabular}{|cccc|}
			\hline
			\multicolumn{2}{|c}{LaOBiS$_{2}$} & \multicolumn{2}{c|}{P4mm (\#99)} \\ \hline
			& \textit{a}(\AA{})            & 4.0544           &                 \\
			& \textit{b}(\AA{})            & 13.8246          &                 \\
			& & & \\
			& x & y & z\\
		La           & 0            & 0             & 0.0904          \\
		La           & 0.5          & 0.5           & 0.9096          \\
		Bi           & 0            & 0             & 0.6312          \\
		Bi           & 0.5          & 0.5           & 0.3688          \\
		S1           & 0            & 0             & 0.3832          \\
		S1           & 0.5          & 0.5           & 0.6168          \\
		S2           & 0            & 0             & 0.8096          \\
		S2           & 0.5          & 0.5           & 0.1904          \\
		O            & 0.5          & 0             & 0       \\
		    \hline   
	\end{tabular}
\end{table}

\begin{table}[ht]
	\centering
	\caption{The tested model of P\={4}m2 [8] symmetry}
	\label{Table S4}

		\begin{tabular}{|cccc|}
			\hline
			\multicolumn{2}{|c}{LaOBiS$_{2}$} & \multicolumn{2}{c|}{P\={4}m2 (\#115)} \\ \hline
			& \textit{a}(\AA{})            & 4.1091           &                 \\
			& \textit{b}(\AA{})            & 13.4196          &                 \\
			& & & \\
			& x & y & z\\
			
		La           & 0.5          & 0             & 0.1073           \\
		Bi           & 0            & 0.5           & 0.3855           \\
		S1           & 0.5          & 0             & 0.3844           \\
		S2           & 0.5          & 0             & 0.8128           \\
		O            & 0            & 0             & 0                \\
		O            & 0.5          & 0.5           & 0          \\
		\hline     
	\end{tabular}
\end{table}

\begin{table}[]
	\centering
	\caption{The tested model of P4/nmm [26] symmetry}
	\label{Table S5}
\begin{tabular}{|cccc|}
	\hline
	\multicolumn{2}{|c}{LaOBiS$_{2}$} & \multicolumn{2}{c|}{P4/nmm (\#129)} \\ \hline
	& \textit{a}(\AA{})            & 4.0544          &                 \\
	& \textit{b}(\AA{})            & 13.8246          &                 \\
	& & & \\
	& x & y & z\\
			
		La           & 0.5          & 0              & 0.0904            \\
		Bi           & 0            & 0.5            & 0.3688            \\
		S1           & 0.5          & 0              & 0.3832            \\
		S2           & 0.5          & 0              & 0.8096            \\
		O            & 0            & 0              & 0   \\
		\hline             
	\end{tabular}
\end{table}
	
\end{document}